# Pore space analysis of beech wood – the vessel network

**P. Hass*[1], F.K. Wittel[1], S. A. McDonald[2,3], F. Marone[2], M. Stampanoni[2,4], H.J. Herrmann[1], and P. Niemz[1]**

[1] Institute for Building Materials, ETH Zurich, Schafmattstrasse 6, CH-8093 Zurich. *phass@ethz.ch

[2] Swiss Light Source, Paul Scherrer Institut, 5232 Villigen, Switzerland

[3] Department of Radiology, University of Lausanne Medical School, Lausanne, Switzerland, now at University of Manchester, School of Materials, Manchester.

[4] Institute for Biomedical Engineering, University and ETH Zürich, 8092 Zürich, Switzerland

## Abstract

Water transport in wood is vital for the survival of trees. With synchrotron radiation X-ray tomographic microscopy (SRXTM), it becomes possible to characterize and quantify the 3D network formed by vessels that are responsible for longitudinal transport. In the present paper, the spatial size dependence of vessels and the organization inside single growth rings in terms of vessel induced porosity was studied by SRXTM. Network characteristics, such as connectivity, were deduced by digital image analysis from the processed tomographic data and related to known complex network topologies.

## Introduction

Knowledge of the anatomical structure of wood is crucial towards understanding various biological processes in trees and physical properties of wood. To describe transport processes, for example, a characterization of the three dimensional (3D) wood structure is needed. It is still not fully understood how a tree is able to transport water from the roots to the crown at a height of more than 100 m (Koch et al. 2004; Koch and Sillett 2009; Netting 2009). It has also since been realized that the cohesion tension theory, proposed in the late 19[th] century, does not sufficiently explain water transport via the capillary network in the xylem (Becker et al. 2000; Meinzer et al. 2001; Hölttä 2005). The structure of the wood also has a strong impact on its technological applications, such as bonding with adhesives, impregnation with preservatives, or coating (Kamke and Lee 2007).

With the advent of microscopy in the 17[th] century began the description of basic components of wood like fibers, vessels, wood rays, and pits between the cells (Bosshard and Kučera 1973; Wagenführ 1999). In the 19[th] century, bordered pits and cell wall layers became visible (Preston 1965; Côte 1967; Carlquist 1988). In the 20[th] century, X-ray diffraction or electron microscopy led to a profound knowledge of the cellular ultrastructure. For example, the crystalline structure of the cellulose, the fine pits in the cell walls, and the cell wall structure could be studied with high resolution (Harada 1962; Preston 1965; Côte 1967; Carlquist 1988; Wagenführ 1999). Chaffey (2002) gives an overview on this topic.

Pits are interconnections between tracheids and vessels and permit fluid transportation between water conducting structure elements. The spatial organization of this 3D network determines the functionality and robustness of water transport. The analytical power of 2D observations in this field is limited. The 3D vessel structure became accessible via the "optical shuttle method" of Zimmermann and Tomlinson (1966, 1967). Bosshard and Kučera (1973) went on to investigate the 3D spatial distribution of the vessel network in beech (*Fagus sylvatica* L.). They found that vessels are deflected mainly in the tangential direction from their longitudinal orientation and that they build a fully connected network with a mean contact length of 1.75 mm. With a similar method, Fujita and Sakai (1996) analyzed the vessels in Japanese horse chestnut (*Aesculus turbinata*). Butterfield (1993) found that recordings of individual cross sections proved to be the best means of observing the vessel pathway. Modern computer technology and digital image analysis permits the conversion of 2D images to 3D models (e.g. Bardage and Daniel 2004). Today, spatial investigations of wood are even easier by means of modern micro-computed tomography (μ-CT). For example, Steppe et al. (2004) analyzed vessel characteristics of two wood species by desktop X-ray μ-CT. These authors demonstrated the high potential of a computer-assisted evaluation based on image analysis. Mannes et al. (2009) showed that spatial investigation of anatomical features is possible with synchrotron radiation with a resolution of a few μm. Bucur (2003) also gave an overview on such non-destructive methods.

To quantify the pore space of wood with high resolution, modern μ-CT are the methods of choice. In combination with image processing and image analysis it is possible to determine complete network characteristics beyond single vessel properties. The present work aims at the collection of high resolution SRXTM data from beech wood and their evaluation as 3D images.

## Material and methods

Clear beech wood samples, without any red heartwood, were investigated using SRXTM methodology at the TOMCAT beamline (Stampanoni et al. 2006) at the Swiss Light Source (PSI Villigen Switzerland), following the experimental setup by Mannes et al. (2009). A cylindrical shaped specimen was chosen to reduce the risk of artifacts during reconstruction. Samples were turned with a lathe chisel until a diameter of 3 mm along the longitudinal axis and were kept at ambient room climate (ca. 20°C and 40-50% RH) to minimize dimensional changes. Before turning, the specimens were assembled by bonding two pieces of wood, since the same set of samples will be used for an investigation of the penetration behavior of adhesives into beech wood. Thus, the investigated growth ring sections do not range over the whole sample diameter and the samples contain a bond line in the middle of the specimen. For each sample, a data set of 1501 projections over 180º was acquired at a photon energy of 10 keV. The X-rays were converted to visible light by a 25 μm thick Eu doped LAG scintillator, magnified (4x) by an optical microscope and detected by a 14 bit CCD camera (2048 x 2048 chip). Two fold on-chip binning was used, giving a nominal pixel size of 3.7 μm.

The tomographic dataset was reconstructed based on a Filtered Back Projection algorithm. Finally, the tomographic data were converted to a stack of axial slices (i.e. cross section images), representing layers with a thickness equal to a voxel size of 3.7 μm (Figure 1a).

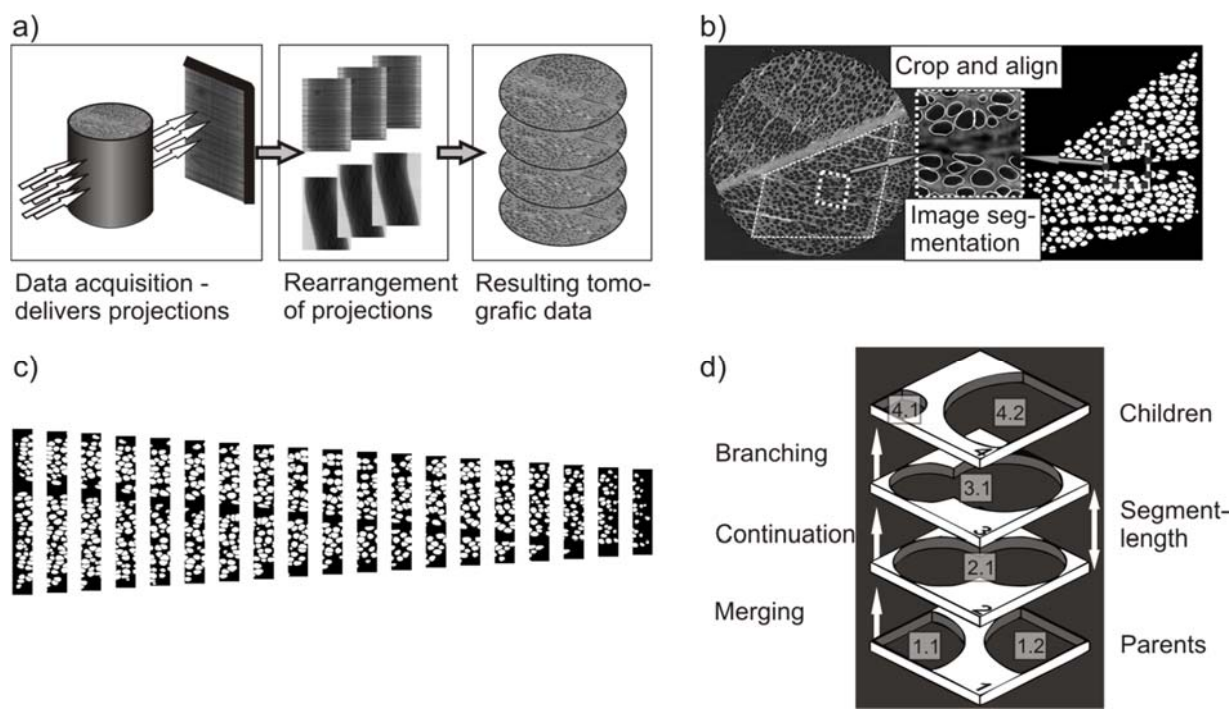

**Figure 1:** Schematic workflow (a) and details of the experiments (b-d). b) Transformation of the selected growth ring in the original cross section to a binary image; the subset demonstrates the quality of the segmentation method. c) Binning of a growth ring for porosity measurements. d) Principle of connectivity analysis.

A series of image enhancement and processing operations were performed: First, the gray scale values were adjusted to increase the contrast in all individual cross sections. For data reduction, single growth ring sections were isolated. The selected regions were rotated to align all samples parallel to the three major wood directions (x = tangential; y = radial; z = longitudinal). To increase the precision of the image segmentation, the resolution of the images was doubled, before a two-step image segmentation was applied.

The first step was based on local gray value thresholds of certain regions. This step already delivers reasonable results since the gray values of the wood substance and the air in the pores differ distinctly. Although, some regions are incorrectly segmented, since the cells in the wood rays are also identified along with the vessels. However since these regions are significantly smaller than the vessels and as they do not extend over multiple layers, they can be eliminated easily by morphological image processing steps of the binarized data. In our case, a morphological opening with a circular structuring element was used. At the end of the process, the pores of the selected growth ring were obtained (Figure 1b).

To analyze the vessel network, a so called layer-wise blob analysis was performed. A blob analysis is an image analysis technique that is used to identify connected regions in a binarized image. First, all white areas in the cross sections were labeled and their areas, equivalent diameters, and positions were stored. By normalizing the radial position of the vessels by the growth ring width, all values were expressed in relative coordinates for comparison. To measure the relative porosity, the selected area was binned (Figure 1c). The porosity of each bin equals the ratio of white pixels to black ones.

In a second step, those planes had to be identified, where the vessels merge or branch. Note that merging of vessels does not necessarily mean that two vessels become a single one. In many cases, merging means that vessels only touch each other. Pits are visible in cross sections as channels that connect two vessels. In reality, pits have a diameter of 4-11 µm (e. g. Harada 1962; Wagenführ 1999). Note that the resolution of our measurement is not high enough to resolve pits accurately. The gray value of a voxel represents the average value of the attenuation coefficient in a 3.7 x 3.7 x 3.7 µm$^3$ volume. Therefore regions containing pits are closer to the gray value distribution of the pore space and are also segmented.

The process is further demonstrated in Figure 1d: The labeled, adjacent layers were projected onto each other and the related pores are identified. For example, the pore 2.1 in layer 2 is projected onto layer 1. Since this pore is directly linked to pores 1.1 and 1.2, those pores must have merged in layer 2. The projection of layer 3 onto layer 2 only matches one pore (3.1 matches 2.1), since the vessel continues. An example for branching occurs when layer 4 is projected onto layer 3, since pores 4.1 and 4.2 match the single pore 3.1 in layer 3. Pores without counterparts on neighboring layers mark either starting or ending vessels. The number of layers in between determines the segment length. For example in Figure 1d, the segment length would be 2, as it reaches across two layers. The pores, from which a segment emerges, are called parents, while the branching ones are named children. With such a family tree structure, the connectivity of the whole network can be identified.

## Results and discussion

**Pore size and porosity:** The porosity arising from vessels and their characteristic size was analyzed in individual segmented cross sections with a longitudinal distance of 300 µm. Due to the finite length of the samples (3mm), the results can be considered as being invariant in the longitudinal direction. In order to compare the data from the 16 evaluated growth rings with different widths (between 728 µm and 2022 µm), the growth ring width (GRW) was normalized. Thereby, the GRW can be described using dimensionless coordinates ranging from zero for earlywood to one for latewood. In Figure 2, the average values for each sample in the respective region of the growth ring is shown (dots in Figure 2).

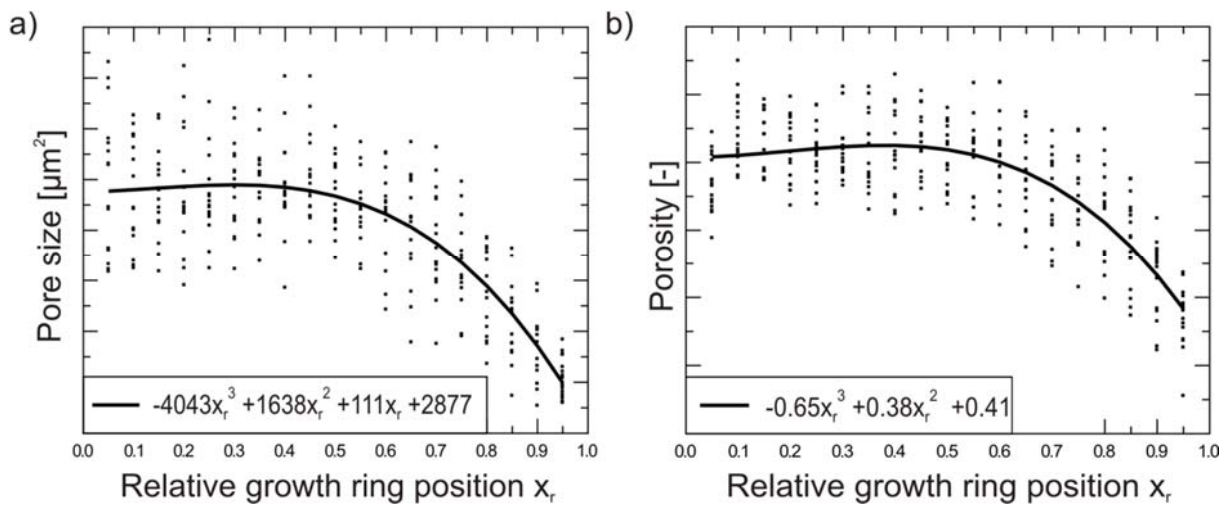

**Figure 2:** Pore sizes (a) and porosities (b) as a function of a dimensionless relative position in the growth ring (in which: 0.0 represents earlywood and 1.0 represents latewood). The polynomial equations describe the correlation curves presented.

The average vessel diameter of beech should be less than 50 µm (Wagenführ 1999). However, the vessel diameters of all investigated growth rings gave a mean value of 55.3 ± 11.7 µm. Of course, pore sizes greatly vary, depending on their radial position.

The high variance in vessel size and the differences in GRW are partly due to environmental factors (Sass and Eckstein 1995; Pumijumnong et al. 2000). The samples in the present work were selected randomly, thus the pore size variation reflects the natural scattering of this anatomical feature. Essentially, the results of the present study are in agreement with those of Sass and Eckstein (1995), who investigated the vessel size distribution of one growth ring of beech wood from the ecophysiological approach.

Additionally, values for pore size and porosity depending on growth ring position are fitted using a cubic polynomial function. As expected for a diffuse-porous wood such as beech, no abrupt change within one growth ring from earlywood to latewood can be observed. The functional form of those fits is quite similar for all specimens; they differ mainly in their individual axis intercepts. By averaging all fits, it is possible to describe the relation between relative position inside the growth ring and pore size or porosity with one final expression (solid line in Figure 2). In the end, it illustrates that, rather than the GRW, it is the relative position inside the growth ring that determines the mean pore size and the porosity. Hence the construction plan of a beech tree seems to follow the principle of keeping relative pore sizes and porosities more or less constant, which makes sense in terms of keeping the water transport distribution constant. Only the GRW is adapted as a reaction to the yearly changing growth conditions (Sass and Eckstein 1995; Pumijumnong et al. 2000).

**Connectivity and network topology:** Water transport to the crown is mainly influenced by pore sizes and capillary forces, as well as porosity influence on the flow rates. Additionally, the way the vessels are interconnected is also crucial. The reconstructed vessel network shows several important characteristics, such as missing vessel contacts across growth ring borders (Figure 3a) or the weaving around of wood rays (Figure 3b).

As a consequence of the deflection around rays, vessels contact each other more frequently in the tangential, rather than in the radial direction. At contact zones, transport between vessels takes place through pits, leading to a homogeneous distribution of water throughout the entire growth ring. This tangential orientation of the vessels is common in all tree species and allows the tangential spreading of the water throughout the whole growth ring during the transport in the axial direction. Additionally, the vessels develop helically around the stem (Zimmermann 1983).

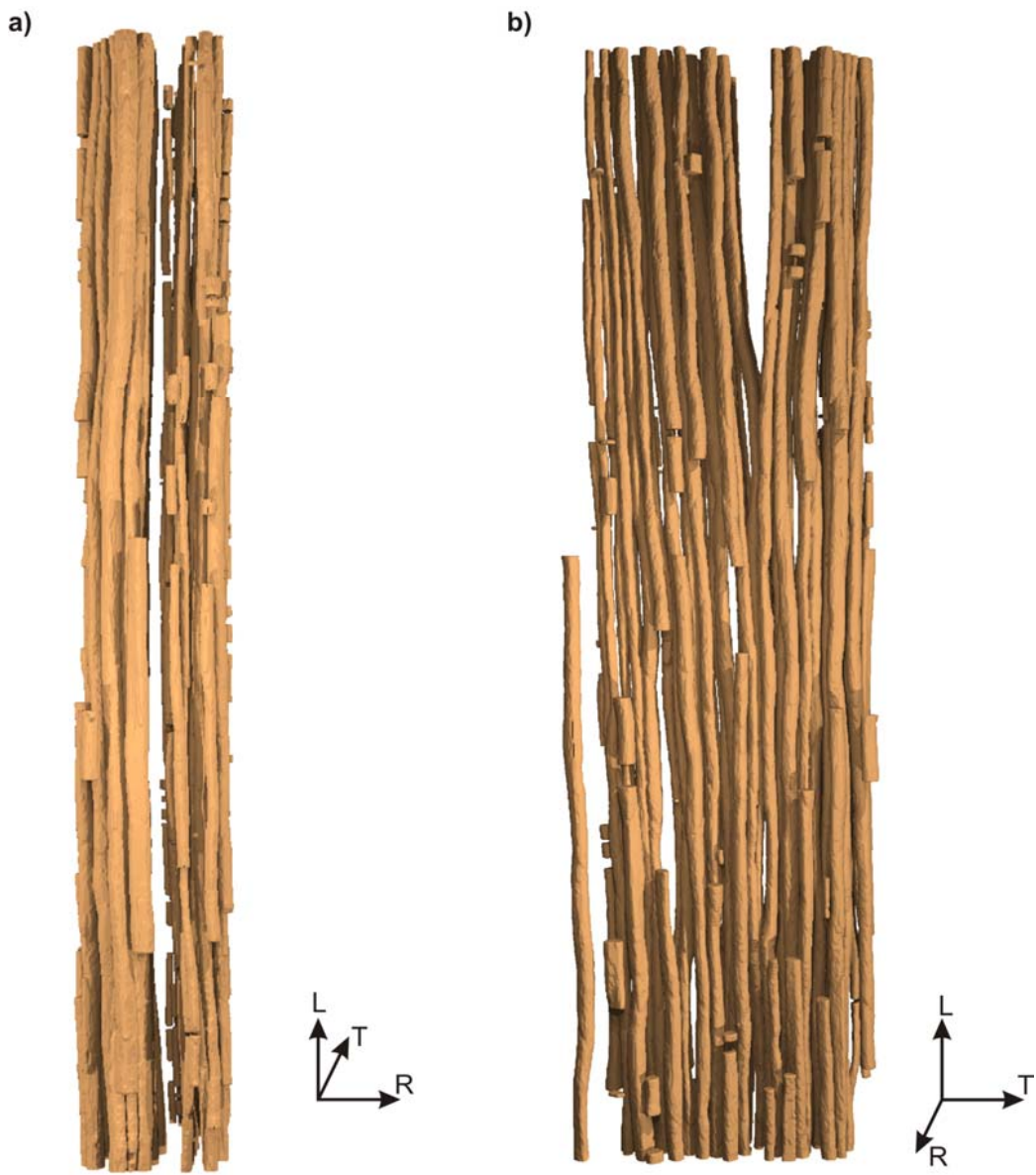

**Figure 3** Vessel network of beech wood looking along T-direction (a) and R-direction (b). Sample height is 3 mm. Note that the displayed vessel discontinuities are mainly due to vessels moving in and out of the extracted volume.

To characterize the vessel network, connectivity analysis was performed as described in "Material and methods" of this article (Figure 1d). Each segment is labeled, thus the first and final layer is obtained, as well as the labels of parent and children segments if present. Knowing the connectivity, a cluster analysis is performed to identify interconnected vessel segments (cluster). Note that the cluster number depends on the size of the sample. Unlike softwoods and some ring porous hardwoods, beech builds up a vessel network in each growth ring without any vessel contacts across growth ring borders (Bosshard 1976). The transport in the radial direction over growth ring boundaries is managed by wood rays (Kučera 1975). Therefore, the analysis of a complete growth ring would have resulted in one single cluster, presuming the system has not been damaged. As the samples used in the present study depict only a small section of the growth ring, numerous clusters with differing sizes can be identified. However, the orientation of clusters is evidently in the tangential direction as a consequence of the waviness of the vessels (Figure 4).

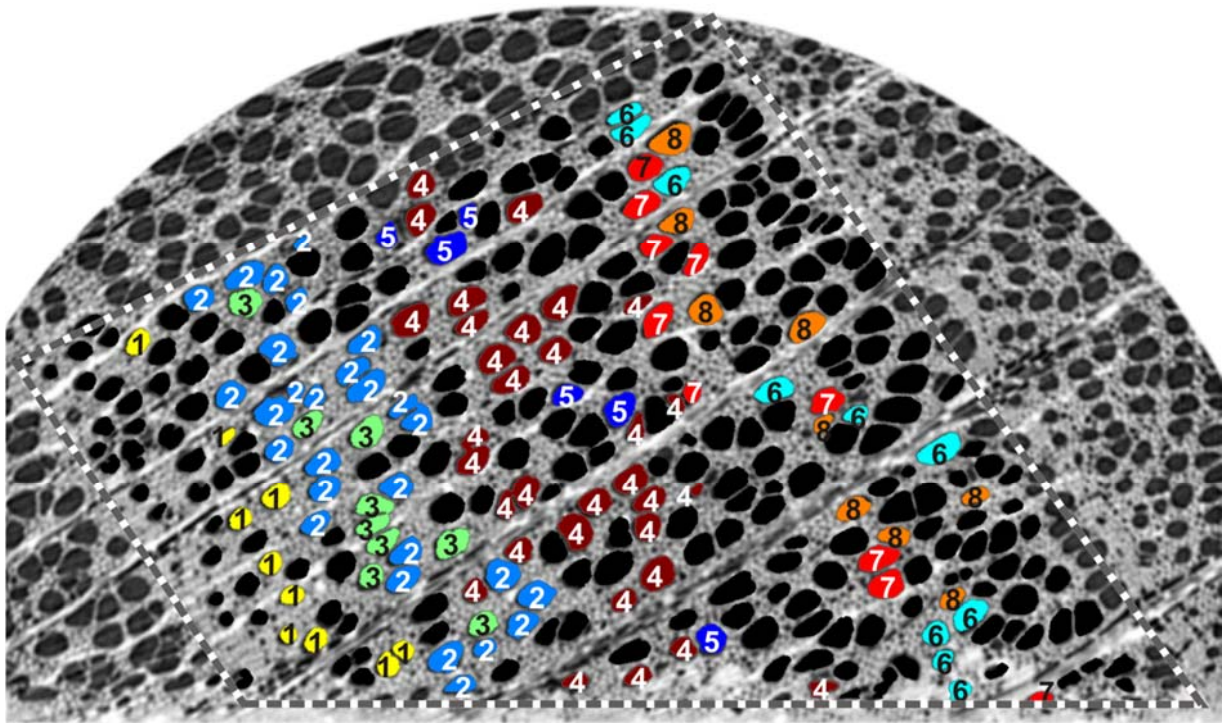

**Figure 4** Projection of the 8 largest clusters in one exemplary cross section. Cluster numbers are printed for clarity.

By setting boundary vessels, it is also possible to calculate the invasion paths from the outside into the network. These paths describe, how a fluid would penetrate the vessel network and where the fluid would be after a given time. This will be very useful for the investigation of the penetration behavior of adhesives that will be conducted on the same set of samples in a later study.

Transporting water inside a network can ensure transport even in the case of partial damage to the network. Dujesiefken et al. (1989) and Schmitt and Liese (1995) investigated the wound reactions of beech wood following an injury and showed how the tree encapsulates the affected area to protect itself from the inflow of air and possible attack by fungi. Zimmermann (1983) reports that trees are even able to maintain water transport when a large part of the network is destroyed, as has been demonstrated in the "double saw-cut experiment", where the stem of a tree was cut halfway across the trunk from opposite sides at different heights.

However, robustness depends on the topology, extent, and type of damage. The allocation of segment lengths allows for characterization of the network topology. After the connectivity analysis described in "Material and methods", the number of layers for each segment, and therefore its segment length, is known. Figure 5 shows the segment length distribution of the identified segments in all samples in double logarithmic scale. It is astonishing that no evidence for a characteristic length of segment can be found; rather two distinct regimes are observed that can be described by power laws with two different exponents. The authors suppose that the exponent for small segment lengths arises from the pit distances inside vessel contact zones with distances up to about 120 µm and the second exponent describes the inter-contact zones that spans over more than one order of magnitude (Figure 5).

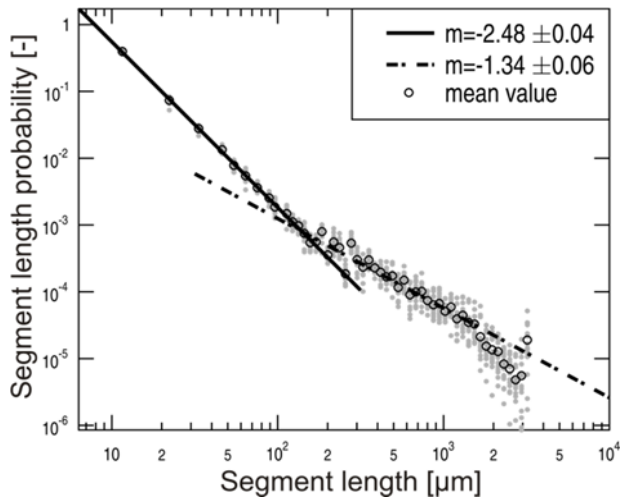

**Figure 5:** Log-log plot of the normalized distribution of segment length probability.

The decline on the right hand side results from the finite size of the sample and is not considered a regime in itself (size effects). Accordingly, characteristics of a scale free network are observed. Those networks are known to be quite robust against random attacks (Albert et al. 2000).

As it is vital for trees to keep up water supply even when damage is done to the transport system, a scale-free network seems to be the best solution: This combines transport in the radial direction via rays and longitudinally through the vessel network. Due to a high connectivity of vessels in the tangential direction, the entire growth ring is involved in the transport from the root to the crown (Bosshard 1984). By building a network with scale-free characteristics, trees have found a way to protect the transport of water against random attacks.

## Conclusions

The 3D pore space in wood can be quantified in a straight forward way by means of SRXTM data combined with advanced image analysis. Due to the automatized routine evaluation, an investigation on large data sets is possible, which strengthen the information value. Important findings are:

- Pore size as well as porosity are similar as a function of the relative position inside one growth ring in the radial direction and can be fitted by a cubic polynomial function.

- The well known tangential vessel orientation was confirmed and is enforced by wood rays, which induce the vessels to deviate from the strong axial direction and weave around the rays. Hence the vessel network shows a greater connectivity in the tangential rather than the radial direction.

In the segment length distribution, no characteristic length was recognized. Furthermore, two distinct regions can be observed that can be expressed via power laws with two different exponents. By building such a network, which presents the characteristics of a scale-free network, beech trees protect their transportation system in a very robust way against random attacks. Further studies could be dedicated to the question whether scale-free networks are a common strategy in woody plants, especially in species with vessel systems that are not isolated in growth rings. The technical application of the findings on the vessel network topology, e.g. on the penetration of fluids into wood, is a subject of further studies in our laboratory.

## Acknowledgements

The financial support of this work under SNF grant 116052 as well as the advice of Anders Kästner (NIAG/ASQ, PSI, Villigen) for the image processing are gratefully acknowledged.